Title: Proposal of Experimental Test of General Relativity Theory

Authors:            Anatoli Vankov
(Physics Department, Eastern Illinois University, cfaav@eiu.edu)

Comments: 28 pages

On the basis of the relativistic mass-energy concept we found that a proper mass of a test particle in a gravitational field depends on a potential energy, hence, a freely falling particle has a varying proper mass. Consequently, a multitude of freely falling reference frames cannot be regarded as a multitude of equivalent inertial reference frames. There is a class of experiments, in which an inner observer can distinguish between the state of free fall in a gravitational field and the state of free space by detecting the effect of a proper mass variation. If so, a demonstration of a violation of the Equivalence Principle is possible. It is shown that a variant of the classical Pound-Rebka-Snider experiment on a photon frequency shift in a gravitational field, if conducted in a freely falling laboratory, would be such a test.

Abbreviation:  SRT- the Special Relativity Theory, GRT- the General Relativity Theory, EP  - the Equivalence Principle, PRS - the Pound-Rebka, Snider (experiment)

### Introduction

The phenomenon of gravity is known to be incompatible with the Special Relativity Theory (SRT), so the General Relativity Theory (GRT) is currently considered the only reasonable field theory of gravity. However, attempts to develop a Quantum Relativistic Gravitational Theory failed. A general feeling about such a situation may be expressed as an expectation of a breakthrough by a discovery of a radically new fundamental physical principle governing gravitational phenomena under both relativistic and quantum-mechanical conditions. Then known problems in the Standard Models of Particle Physics and Cosmology would possibly be understood.

It is thought less probable but not to be excluded that the breakthrough could happen by finding out that the GRT is inherently inconsistent. Any evidence of a violation of the Equivalence Principle would be a signal of such inconsistency, that is, an inherent contradiction. In other words, the GRT being a falsifiable theory could be discarded in principle by a properly formulated observational test revealing the contradiction. In reality a situation is not as simple as that because effects of such significance are usually very small and hard to observe with the required precision. Besides, plenty of room is left for model corrections. Such difficulties are seen, for example, in NSF gravitational experimentation programs.

Among different tests of the GRT [1] one "classical test" has been carried out of weighing a photon in the gravitational field (R.W.Pound, G.A.Rebka, J.L.Snider, the PRS experiment, for abbreviation [2]). Though the measured effect is extremely small the experiment has been eventually conducted with a high precision (about one percent in the effect), and the result was in a good agreement with the GRT prediction. By that time the Equivalence Principle (EP) was firmly justified in different measurements, and the PRS experiment was considered as additional evidence supporting the EP. The GRT predicts the observed photon frequency shift proportional to the strength of a gravitational field. According to the EP the shift disappears as the field "locally disappears" in a freely falling frame carrying both a photon resonance emitter and a corresponding detector. Though the experiment was not reproduced under free fall conditions due to obvious technical complications there is a strong belief among the GRT community that it would result in a zero shift. But we doubt such prediction for reasons explained below. We expect that a violation of the EP will be observed in the PRS experiment if conducted in a freely falling frame. We show that this particular experiment has a status of a new independent test of the EP, and propose to conduct it.

### *1. Current status of the GRT*

The EP is considered the physical basis of the GRT. But some GRT experts argue [3,4] that there is no need to refer to the EP while formulating the GRT because the EP mathematically reflects a trivial fact of possibility to locally approximate a curved 4-space by a flat space metric. However, the physical meaning of the EP seems to go beyond the mathematical treatment. The EP physical formulation is as follows. *In an arbitrary gravitational field no local experiment can distinguish a freely falling non-rotating system (local inertial frame) from a uniformly moving system in the absence of a gravitational field.* There is another formulation of the EP: *at every point of space-time it is possible to choose local coordinates so that all physical laws take the same form as they are formulated in the SRT in the absence of gravity.*

It was shown (see, for example [5]) that the EP rests on the equality of the gravitational and inertial mass, hence, the equality is postulated in rather than deduced from the GRT. At the same time the GRT predicts some *Machian effects* (frame dragging, in particular) revealing specific properties of inertial mass. Those effects, in fact, show the way of detecting a source of the "disappearing". The problem is that GRT predictions are often subject to theoretical approximations especially concerning momentum-energy exchange between parts of a system. Therefore, "a theory of GRT theories" started emerging for explaining the meaning of what is tested [1]. There was a troubling moment in the GRT history when it was realized that the energy-momentum *pseudo-tensor* could not adequately account for a "locally disappearing" gravitational energy. Though the pseudo-tensor still looks much as a foreign body in the set of tensorial field equations, one may argue that it is not a drawback of the theory but rather nature behaves accordingly.

Among disputable issues of modern Physics the most important one is a failure of the GRT to be subject to renormalization and quantization. It makes the current situation in the gravitational theory quite problematic and very challenging. In this situation we are going to re-analyze the old problem of physical validity of the EP. The starting point will be the relativistic phenomenon of a proper mass variability, which follows from the SRT but ignored in the GRT.

## 2. A proper mass variability in the SRT mass-energy concept

Let us consider a point-like particle of mass $m$ in a central gravitational force field with a potential $\phi(x)$ produced by a solid sphere of mass $M_R >> m$ and radius $R$:

$$\phi(x) = -GM_R / x \qquad (x \geq R) \qquad (1)$$

where $G$ - the gravitational constant, $x$ - the distance between the center of the sphere and the particle. One may consider a model, in which the particle can be slowly moved along a radial direction with use of some ideal transporting device supplied with energy. The whole system is isolated. In accordance with the SRT mass-energy concept a change of potential energy of the particle should be equal to the corresponding change of its proper (rest) mass:

$$dm = [GM\,m(x)/c^2 x^2]\,dx, \quad (x \geq R) \qquad (2)$$

In this conceptual example the question about work-energy balance (does the particle perform work?) is viewed differently in the SRT than in the GRT. In the latter a proper mass is assumed to be constant while a field provides the particle with energy ("the field works on the particle"). Then the controversy arises about a separation of energy of a field and a total mass-energy of the system. In the SRT there is no room for the field energy additional to the total mass-energy of interacting objects. Remember, in the static case of a conservative force field a transporting (supporting) device with an energy source should be included in the system. Due to an energy exchange between the particle and the transporting device the particle turns to be bound. The total energy of the particle under consideration is characterized by two varying components: the proper mass-energy and the binding energy, the sum of them being constant.

The above difference between the SRT and the GRT is of principle importance. Both theories claim to be competent to consider the conceptual example. They look apparently self-consistent but result in different Physics. Bearing this warning in mind we will continue to consider the conceptual example on the SRT basis.

From (2) the proper mass is found as a function of the distance:

$$m(x) = m\exp(-x_R/x), \quad (x \geq R) \tag{3}$$

or in a "weak field" approximation:

$$m(x) = m(1 - x_R/x), \quad (x >> x_R) \tag{3a}$$

where $x_R = GM/c^2$. As is seen from (2) and (3), the proper mass of the particle has to be reduced by the equivalent amount of energy to be given to the transporting device. A binding energy (a mass defect) should be interpreted as a potential energy change. In the general case of a multi-particle isolated system, the total mass defect is shared by interacting particles, an individual share being determined by a particle mass-space distribution. The behavior of the system seems to be governed by the principle: the *proper mass of the system tends to minimum*. Using the same approach one can see, for example, that in the case of two identical point-like gravitating particles a proper mass should be a function of the distance $x$ between them:

$$m(x) = m/(1 + x_0/x) \tag{4}$$

where $m$, as before, is the maximal proper mass at infinity, and $x_0 = Gm/c^2$. In the "weak field" approximation (4) is identical to (3a).

It is clear that the static potential $\phi(x) \sim 1/x$ in fact describes an asymptotic behavior of interacting particles at a large distance comparing with $x_R$ or $x_0$ where a proper mass change is negligible. In the limit as $x \to 0$ in [4] the proper mass vanishes as a result of the "energy exhaustion" effect. When a proper mass variability (4) is taken into account a gravitational force between two identical point-like particles takes a form

$$F(x) = mc^2 x_0/(x + x_0)^2 \tag{5}$$

The "exhaustion" effect means that a singularity of a point-like source

(that is, classical "self-energy" divergence) turns out to be eliminated in agreement with the SRT mass-energy concept. Notice that a "field' in the GRT is "inexhaustible" due to above divergence.

In the case of the potential generated by a solid gravitating sphere (1) a force exerted on a test particle along with its proper mass should vanish on the surface of the sphere with a sufficiently big ratio $M_R/R$ (say, in the gravitational collapse case)

$$F(x) = (mc^2 x_R / x^2)\exp(-x_R/x), \quad (x_R > R, \ x > R) \qquad (6)$$

At $x = x_R$ a force is maximal, therefore, in the region ($R < x < x_R$) the particle is "locked": the force increases with a distance (a confinement effect).

In accordance with (6) the potential energy function in general has a form:

$$m\phi(x) = \int_x^\infty F(x)\,dx = -mc^2[1 - \exp(-x_R/x)], \quad (x \geq R) \qquad (6a)$$

where $m$ is the proper mass at infinity, and $x_R = GM_R/c^2$, as in (3). In (6) and (6a) we have a relativistic form of a classical gravitational force and potential function for a massive sphere with the Newtonian limit at $x \gg x_R$.

The phenomenon of a proper mass variability being a consequence of the general SRT mass-energy concept takes place in any type of interaction. Nuclear energy is the commonly known example. Let us consider the effect in the case of the Coulomb force. Assuming that an electric charge is not affected by a proper mass variation (in accordance with observations) one can find the effect due to particle-antiparticle electric attractive force:

$$dm = (kq^2/c^2 x^2)dx \qquad (7)$$

where $k$ is the electric constant in the Coulomb law, and $q$ is the electric charge. Then

$$m(x) = m(1 - x_a/x), \quad (x \geq x_a) \qquad (8)$$

A proper mass vanishes at $x_a = kq^2/mc^2$ (classical electromagnetic radius). It means that at this distance "something happens", apparently, annihilation when Coulomb potential energy turns to be equal to the proper mass energy equivalent. This is a remarkable result because it shows that both the gravitational and the Coulomb potential have the same source of energy, a proper mass. One has to conclude, for example, that the energy stored in a charged capacitor is due to its increased proper mass as a result of work performed by an external inertial force in a process of charging (a charged sphere is heavier). Obviously, in the case of a repulsive force between charge-like particles a proper mass change is opposite to that for unlike charge particles. The effect is appreciable at distances comparable with $x_a$, and indefinitely increases when $x \to 0$:

$$m(x) = m\,(1 + x_a / x) \qquad (9)$$

Remember that we are restricted to the SRT assumption of point-like particles. Hence the above critical distances so far should be treated as parameters of the corresponding classical model, and should be reinterpreted under quantum-mechanical conditions. Classical gravitational and electromagnetic theories also deal with point-like masses and charges. *The obtained new result of fundamental importance is a connection between the two types of interaction through the proper mass, and singularity elimination due to departure of a potential function from the inverse distance law at small distance.*

In Newtonian Physics the force-energy theorem and the mechanical energy conservation law are formulated without an explanation of the physical meaning of potential and kinetic energy. The GRT is not very helpful in this respect either. The SRT mass-energy concept makes terms defined on the relativistic basis of the total energy conservation law in consistence with Newtonian Physics.

The relative effect of a proper mass variation $dm/m$ is extremely small under Earth or Sun conditions. An order of smallness of $dm/m$ characterizes a scale of energy exchange between interacting parts of a material system in comparison with an ultimate energy stored in a material object under consideration. In attempts to understand a matter-matter interaction through energy exchange one has to study the proper mass variation as a relativistic effect no matter how small it is.

As is seen, a proper mass is constant when an object does not participate in any dynamical process accompanied by energy exchange. Obviously, static potential in a multi-body system may be found in the "weak field" (linear) approximation by integration over sources in the manner of the Coulomb potential. In general, *the classical gravitational field turns out to be non-linear through a proper mass dependence on potential energy,* a gravitational field concept in Newtonian Physics being a "weak field" limit. The reason for non-linearity in this approach is quite different from that in the GRT.

At this point we have to state that a classical electromagnetic field theory, which is commonly considered a principally linear field theory, in fact, should be generalized in the same way. A correspondingly formulated relativistic field theory should operate with charged particles of varying proper mass in the presence of inertial and "electromotive" forces. This approach seems to be a basis for a unification of classical gravitational and electromagnetic forces and completion of a consistent theory of radiation of accelerated particles.

Further we use the term "gravitational field" in a formal sense of a gravitational force field. As was mentioned in *Introduction*, the SRT Dynamics equations of motion of a particle in a gravitational field were allegedly found contradictory to observations. This controversial situation should be investigated in view of the above results.

### 3. *Is the SRT really inconsistent with gravitational observations?*

The GRT idea of "a field as a carrier of a gravitational energy, which is a source of a secondary field, etc" is contradictory to the SRT Mechanics. In reverse, the SRT mass-energy concept is incompatible with the GRT, as was previously shown. There might be different illustrations of the situation.

a. Consider annihilation of a pair of a slow neutron and an antineutron, or any particle-antiparticle pair. Observation of this process shows the exact SRT mass-energy balance with no energy gained *from the field* in addition to the energy equivalent of the initial proper mass. Nature does not reveal a divergence in "self-energy". The SRT divergence free concept is totally consistent with this fact.

   b. According to the GRT a kinetic mass is the one of a gravitating type, hence, an assessment of the basic cosmological parameter (the critical mass density, which determines a universe gravity pull) should include a kinetic mass of both massive matter and massless particles. In reality, the assessment is based on counting a proper mass only. (For this reason the neutrino mass problem is one of "hot topics" in cosmology). Cosmologists seem to ignore the GRT prescription on this account being aware that kinetic mass-energy depends on a reference frame choice, therefore, it should be separated from gravitating matter.

   c. The GRT predicts gravitational and electromagnetic radiation of an accelerating particle or a body, in particular, in a free fall state. For example, a gravitational radiation from binary stars is searched by means of energy-momentum balance counting. Both stars are in free fall in a local gravitational field. There is no reason why a local observer would not be able to detect the radiation in the same manner as an earth observer. Therefore, had the radiation been unambiguously found it could have manifested a violation of the EP. The same argument arises from a thought experiment with a charged particle emitting a radiation in free fall.

   There were attempts to develop a gravity field in a flat (Minkowski) space [6]. The conclusion was made that the SRT is not compatible with gravity. For example, a scalar field was ruled out because it does not couple gravity to light. As was emphasized, in the GRT a freely falling particle gains energy from the field, and the same is assumed true for a photon. Therefore, gravitational properties of a photon should be the problem one has especially to be concerned with when comparing the theories. An apparently strong GRT position in this issue is a statement that *all forms of energy are subject to gravitation,* electromagnetic waves (light) included. The statement in fact has a status of a postulate in addition to the EP. Both the GRT and the SRT, having no quantum-mechanical extension in a gravity theory, cannot provide a theoretical basis for gravitational properties of massless particles. In practice both are guided by circumstantial arguments.

In contrast to the GRT the SRT gives a picture of a freely falling particle, which spends its own source of energy (a proper mass) when gaining kinetic energy. A physical nature of the kinetic energy is not clear at this level. Evidently, we need a quantum-mechanical interpretation of an interaction of a particle with "physical vacuum". Further we will see that regardless of the quantum-mechanical properties of the kinetic mass its dynamical role in the presence of gravitational and inertial forces can be well understood in the SRT. As was discussed above, one should expect that the kinetic mass does not gravitate. Similarly, a photon having no proper mass is not expected to change its energy in a gravitational field. Then how to interpret the PRS experiment, which showed a photon frequency shift?

According to the GRT the shift is due to a change of energy (frequency) of a *gravitating photon* in a gravitational field, so the PRS experiment was treated on this basis. Consequently, the shift would disappear if measured in free fall because the field disappears. We suggest another picture. A *non-gravitating photon* does not change its frequency in flight. This is a difference in resonant frequency of an emitter and a detector placed at different equipotential surfaces that causes the observed effect. In other words, the resonant shift is due to a difference in a proper mass of identical nuclei of the emitter and the detector. Which interpretation of the experiment is true may be and should be verified in some crucial test. In the following section we will discuss this issue in detail and show that the *PRS experiment in a freely falling frame* would be such a test. We predict that in this variant a measured frequency shift would be the same as in the original PRS stationary experiment. A confirmation of this statement would mean a violation of the EP. So, the proposed test is vital in the problem of compatibility of the SRT with a gravitational theory. The problem is considered furhter in more detail.

### *4. Brief review of the SRT Mechanics*

The SRT Mechanics is generally known as the Relativistic Kinematics of a free particle considered in some "inertial" reference frame. To deal with a process of transition from one inertial frame to another we need to use the SRT dynamical equations, which are invariant under Lorentz transformation:

$$\frac{d}{d\tau}[m(\tau)\frac{dx^\alpha}{d\tau}] = K^\alpha, \quad (\alpha = 1,2,3,4) \tag{10}$$

They describe a particle motion on some world line $x^\alpha(\tau)$ in an inertial reference frame with Minkowski coordinates $x^\alpha$ and a 4-velocity $\frac{dx^\alpha}{d\tau}$, where $K^\alpha(\tau)$ is a Minkowski 4-force vector, and $d\tau$ is a line arc-length. The equations (10) have been found as a generalization of Newtonian Mechanics by means of the Lagrangian formulation of Relativistic Mechanics. They show that the rate of 4-momentum change equals the Minkovski force. By definition of a time-like world line of a moving particle we have the fifth equation:

$$\frac{dx^\alpha}{d\tau}\frac{dx^\alpha}{d\tau} = -1 \tag{11}$$

which makes the problem definite with respect to five unknown functions $x^\alpha(\tau)$, $m(\tau)$. As was emphasized by Synge [3], a proper mass variation along the world line is explicitly seen from the next equation obtained from (10) and (11):

$$\frac{dm}{d\tau}\frac{dx^\alpha}{d\tau} + m\frac{d^2x^\alpha}{d\tau^2} = K^\alpha \tag{12}$$

A photon is a massless particle, hence for a photon $\frac{dx^\alpha}{d\tau}\frac{dx^\alpha}{d\tau} = 0$ in a flat space having the Minkowski metric $\eta_{\alpha\beta}$

$$d\tau^2 = -\eta_{\alpha\beta}dx^\alpha dx^\beta = -dx^\alpha dx_\alpha \tag{13}$$

A free particle has a constant proper mass, hence, the equation of motion of a free particle is

$$\frac{d^2x^\alpha}{d^2\tau} = 0 \tag{14a}$$

In a non-inertial reference frame one may use a general coordinate system $[x^\alpha(x^\beta)]$. Then equation (14a) becomes

$$\frac{d^2 x^\alpha}{d\tau^2} + \Gamma^\alpha_{\beta\gamma} \frac{dx^\beta}{d\tau} \frac{dx^\gamma}{d\tau} = 0 \tag{14b}$$

where $\Gamma^\alpha_{\beta\gamma}$ is the metric connection of $g_{\alpha\beta}$:

$$d\tau^2 = -g_{\alpha\beta} dx^\alpha dx^\beta \tag{15}$$

So far, nothing has happened: we have the same particle in the same state of free motion, described in an arbitrarily chosen coordinate system. Formally the flat metric $g_{\alpha\beta}$ is not the Minkowski $\eta_{\alpha\beta}$ but this difference is a matter of pure mathematical treatment. A physical interpretation begins with the introduction of the EP in the presence of a force field. The EP requires an appropriate form of $\Gamma^\alpha_{\beta\gamma}$ being a metric connection of a *non-flat metric* $g_{\alpha\beta}$ in the presence of gravitational and inertial forces. *The intrinsic curvature of space-time (materialization of geometry) is now introduced,* consequently, $\Gamma^\alpha_{\beta\gamma}$, and $g_{\alpha\beta}$ should be interpreted in terms of a force field. At this point the GRT has come into conflict with the SRT because a proper mass of a particle in free fall should depend on a gravitational potential to comply with the energy conservation law in its SRT form. Thus, dynamical properties of a relativistic mass are ignored in the EP. Remember that the EP rests on the postulate of the equality of gravitational and inertial mass, but a formulation of the postulate loses a physical sense in the Relativistic Dynamics.

For the sake of convenience and practical usage one may come from the description in space-time ($\alpha = 1,2,3,4$) to the description in 3-space ($i = 1,2,3$) and time $t$ ($\alpha = 4$) using the relation $d\tau = c\,dt/\gamma$ and introducing relative ("ordinary") forces $F_i$:

$$F^i = K^i c^2 / \gamma \tag{16}$$

Then the equations of motion take the form:

$$\frac{d}{dt}(m\gamma u^i) = F^i \tag{17a}$$

$$\frac{d}{dt}(\gamma m c^2) = F^i u_i + \frac{c^2}{\gamma} \frac{dm}{dt} \tag{17b}$$

where $u^i(t) = dx^i / dt$ ($i = 1,2,3$) is the relative 3-velocity, and the proper mass $m$ is time-coordinate dependent. The second equation reflects the total energy conservation law in a relativistic form. Usually it is given in an approximate form, in which a proper mass change in a force field is ignored. The question whether the above equations present a complete form for including a gravitational field will be reopen later in connection with photon gravitational properties. As concerns massive particles, examination of the above equations is interesting in revealing a different role of a proper and kinetic mass in a momentum-energy exchange under action of gravitational and inertial force.

Obviously, any constant force exerted on a massive particle causes a change of momentum but not a proper mass. In correspondence with (10-12) a kinetic mass-energy changes if the force is not perpendicular to the velocity. If it is perpendicular as in a circular motion, the momentum changes direction (but not magnitude) while the kinetic and total mass-energy remain constant and the particle is kept bound. *Its proper "bound" mass is less than that at infinity.* If the particle is in a state of free fall from infinity (initially being at rest) the total mass-energy $m_t$ (the sum of varying both proper $m_p$ and kinetic $m_k$ mass) equals the proper mass $m$ of a free particle at infinity:

$$m_t = m_k(t) + m_p(t) = \gamma(t) m_p(t) = m , \quad (m_p(t) \leq m) \quad (18a)$$

with $\gamma(t)$ taken from (3) as $\gamma(t) = m / m(x(t)) = \exp(x_R / x(t))$.

It would be different in the case of uniform acceleration of a particle due to a constant inertial force. A particle continually gains kinetic energy from an external energy source while *a proper mass $m_p$ is being constant but bigger than $m$* (the mass at infinity) due to gained kinetic mass-energy $m_k$:

$$m_k(t) + m_p = \gamma(t) m_p = m(t) , \quad (m_p \geq m) \quad (18b).$$

Dynamics of this process is obvious. The inertial force has to be initially developed over a time interval $\Delta t_i$ when the force rises. A proper mass rises synchronously, its maximal gain being $\Delta m = m_p - m$, when both the force and the proper mass become constant. During a period of uniform acceleration $\Delta t_u$ the kinetic mass continually increases. A force pulse would be completed when the force drops down to zero over a final time interval $\Delta t_f$. Correspondingly, $\Delta m$ comes to zero, and the particle will be in a state of uniform motion with the mass-energy balance described by the known formula in the SRT Kinematics:

$$m_t = m_k + m = \gamma m \qquad (18c)$$

A *proper mass change* $\Delta m$ *correlates with a force pulse and determines a direction of a mass-energy current between a test particle and its interacting partners. A magnitude of the current is proportional to the proper mass difference, that is, the potential difference developed by a force pulse.* Therefore, a third derivative of coordinate should be generally specified in practical applications of the SRT Dynamics.

A perihelion precession in the Kepler elliptic motion is an example of a problem on a *gravitational* force pulse (therefore, "forbidden" for the SRT). The observation of the effect played a role of one of the seemingly successful tests of the GRT. From the SRT standpoint, the effect is caused by periodic change of a proper mass. A prediction and consistent physical treatment of the phenomenon on the SRT basis could be tried before the GRT had been developed. A nowadays analysis of data raised a lot of questions whether the test has actually led to unambiguous confirmation of the GRT prediction [1]. In our view, the problem should be reconsidered on the SRT basis.

Thus, gravitational and inertial masses are not equivalent in many ways. As was shown, a particle in a state of free fall gains its kinetic mass and a momentum at expense of a proper mass reduction, that is, differently than in an inertial force field. A kinetic mass is differently related to forces of different type either. Consider a gravitational force between two massive particles moving in parallel. The force should be determined by a proper mass of the particles, otherwise, it would depend on an arbitrary choice of a reference frame. The important conclusion drawn from the SRT approach is that the kinetic mass does not gravitate, only the proper mass does.

As is seen from the SRT Dynamics equations, a pulse of any force changes a momentum *by action on a total mass* $\gamma m$. A photon, the total mass of which equals the kinetic one, may be considered as a particle in a limit as $m \to 0$ and $\gamma m = const$. Considering a motion of such a "photon-in-limit" in gravitational field one has to conclude that the result of a force pulse should be a change of momentum but not energy (frequency). Otherwise the equations would not comply with the total energy conservation law stated in the SRT form.

We have to conclude this section with the statement that according to the SRT mass-energy concept a photon in a gravitational field does not change its energy (frequency). In other words, light does not gravitate. The argument that the SRT is not compatible with gravity because of no coupling of light to gravity could be wrong if light does not gravitate in reality. Next we will see that the SRT interpretation of classical tests apparently confirming the GRT picture of a "freely falling photon" may be interpreted in the SRT approach consistently with the concept of non-gravitating photon.

### 5. *The light bending phenomenon*

The light bending phenomenon is directly related to the question of the change of a photon momentum. We need to realize why and how a photon could change direction without changing its total energy (frequency). As was previously discussed, a gravitational force acts on a kinetic mass of a photon. Because of absence of a proper mass a force pulse should result in a change of momentum but not energy (frequency). To keep a frequency constant the photon needs to change a magnitude of velocity when potential changes. Massive particles and photons behave differently in a gravitational field. A massive particle is forced to fall down by gravitational force acting on a proper mass. Consequently, energy and momentum changes. Unlike a massive particle a photon emitted along an equipotential plane in a uniform gravitational field does not change either its direction or speed (it does not fall). Traveling in a radial direction in a central gravitational field the photon changes its speed only. In general, it changes both.

We come to the striking conclusion that the speed of light changes with a gravitational potential, a frequency of the photon in flight being constant. The wavelength changes accordingly. In other words, a light interference with gravity occurs through a change of momentum but not energy. It means that a special form of refraction takes place: photon speed decreases with a decrease of potential energy (and vice versa) with no energy change. The photon behaves in a gravitational field as in a medium with a changing "optical" density as compared to the "free space". Unlike in transparent materials, no reason is seen for dispersion in the gravitational refraction, therefore, one may treat the phenomenon in terms of change of a group speed of light.

A variability of the speed of light is necessary if one requires that gravitational-to-electric force ratio is constant throughout the space. One has to expect that an observer in a freely falling frame cannot identify his state by comparing the two forces. Then the factor of departure from the inverse squared distance law due to the presence of an external gravitational field should be the same for different forces (at least at distances bigger than classical "electromagnetic radius"). If a charge is conserved the electric constant (the pemittivity of space) has to vary in a gravitational field. Consequently, a variation of the speed of light in space around the massive sphere (1) is found as next:

$$c(x)/c = \exp(-x_R/x), \quad (x \geq R) \qquad (19)$$

or, in general, in a field with a potential $\phi(x)$

$$c(x)/c = 1 + \phi(x)/c^2, \quad (x \geq R) \qquad (20)$$

where, as before, $x_R = GM/c^2$, and *c the speed of light in free space*. Therefore, the speed of light becomes a field. Under the "weak field" condition we have

$$c(x)/c \cong 1 - x_R/x, \quad (x > x_R) \qquad (21)$$

and "the index of gravitational refraction" is the inverse quantity. Thus, an outside observer experimenting with light should find a photon slowing down when the photon crosses equipotential surfaces "down" (approaching a gravity center), and speeding up when it travels "up", its frequency being kept constant.

As was emphasized, a pulse of force is needed to change a momentum of a photon by affecting its kinetic (total) mass. The resultant effect depends on a gravitational field configuration and initial condition of a photon emission but not Dynamics of a process. It means that there is no indeed coupling of a photon to gravity. Back to the time before the GRT has been completed the question of variation of the speed of light in a gravitational field was raised by Albert Einstein and others [7]. This idea was shortly abandoned due to its inconsistency with the GRT in its final form, in which a kinetic mass of a photon has the same gravitational properties as any massive particle.

Assessment of the bending angle of light passing the Sun in the SRT approach is consistent with observations, though the same seems true for the GRT assessment. For making a discriminative conclusion more precise measurements are needed with a check of *1/r*-dependence of bending angle and detailed analysis of instrumental uncertainties as well as theoretical assumptions. In the first approximation of the GRT model the angle $\delta$ is given by the formula:

$$\delta = 4M_S G / c^2 R_S \qquad (22)$$

where $M_S$ and $R_S$ are correspondingly the proper mass and the radius of the Sun measured by outside observer at rest. As is known, the angle assessment initially made by Albert Einstein was twice as low as compared to (22). Later the magnitude was doubled to make the assessment consistent with the Schwatzcshild metric characterized by $g_{00} = -(1 - 2GM / c^2 x)$.

There was a discussion on the subject of controversy of the GRT prediction (22) with respect to the EP. This question has a physical sense, because the formula derived purely from the EP is exactly what Albert Einstein initially found (with factor 2 instead of 4 in (22), see, for example, [8]). It shows again that the GRT is based on the EP and the additional field conjecture not well matched (see section 3).

## 6. A frequency shift of a photon in gravitational field

All things considered, our interpretation of the PRS experiment is as follows. For measuring a photon frequency shift both a resonance emitter and a detector should be put at rest at a laboratory frame. Identical nuclei of an emitter and the detector are placed at different equipotential surfaces. Therefore, they have necessarily different proper masses. It is important to use the solid material in both the emitter and detector with the same atomic structure to eliminate a difference in structural binding energy. Then a proper mass difference will be exclusively due to the gravitational potential difference. Thus, a proper mass characterizes energy scaling of matter on an equipotential surface. Obviously, energy of any quantum-mechanical transition is subject to the scaling. This is the cause of a corresponding shift in a resonance frequency of electromagnetic transition in excited nuclei. We have to emphasize again that an emitted photon being in flight does not change its energy (frequency) because its proper mass is zero. If an emitter is placed below a detector the latter will register an apparently red-shifted photon. After linearization of (3) with $g = 9.8 \, m/s^2$ we have a relationship between the change of a proper mass of nuclei $\Delta m$, the distance between equipotential levels $\Delta x$, and the difference in a resonance frequency of the emitter and the detector $\Delta f$ under Earth condition of a uniform gravitational field:

$$\Delta f / f = \Delta m / m = g \, \Delta x / c^2 \qquad (23)$$

where $h$ is the Plank constant.

Now let us discuss the effect of change of the speed of light. In the SRT picture a photon in flight does not change its frequency. Therefore, a wavelength changes. From an outside observer's point of view, a wavelength increases when the photon travels "upward" due to its speeding up. This effect may be called a "red shift". Though a proper frequency of identical ("standard") nuclei depends on a potential, *a wavelength of a "standard" photon in a moment of emission is constant, that is, does not depend on nucleus position.* Standard photons travelling in space carry information on a potential at points of their emission.

Interestingly enough, one can find the same formulas (23) and the same argument of energy conservation law, as concerns a gravitational red shift theory, in textbooks on the General Relativity Theory. But the interpretation is different: $m$ and $\Delta m$ are referred now to the kinetic mass of a photon and its change in a gravitational field. As opposed to the SRT picture, the photon interacts with a gravitational field by energy exchange while a nucleus proper mass is kept constant. The difference in interpretation is essentially physical and can be tested experimentally.

Let us summarize the answer to the question: *what do we expect from the PRS experiment conducted in a freely falling laboratory? In accordance with the SRT mass-energy concept, there is a difference in a proper mass and, correspondingly, in a proper resonance frequency of nuclei of an emitter and a detector placed at different equipotential surfaces. The difference takes place regardless of the reference frame, hence, is not affected by the state of free fall of the laboratory. A photon being emitted by the emitter does not change its frequency in flight. Therefore, in the freely falling laboratory the photon would be registered "red shifted" by a detector in the same way as in the laboratory at rest. It would demonstrate a violation of the EP.*

### *Discussion*

Next we summarize the content of the work and discuss several issues, which are viewed differently in our approach than in the GRT.

### *Importance of the proposed test*

We have found that the SRT interpretation of both the bending light observation and the PRS stationary experiment differs from that in the GRT, though predictions in both approaches are consistent with observations (as far as the effects are of a "weak field" order). We have concluded that a violation of the EP could be observed in certain types of experiments regardless of the field strength. Specifically, this is the PRS experiment, which if conducted in a freely falling laboratory is expected to show the frequency shift the same as under the stationary conditions. The prediction is opposite to that of the GRT (the latter predicts a zero shift). This particular test under free fall conditions has never been performed. Its importance is seen from the fact that the test has a falsifying power: it is intended to demonstrate a violation (if not confirmation) of the EP and distinguish between the two different approaches to the development of gravitational theory. The test is technically realistic because it is similar to the PRS experiment that has already been conducted under stationary conditions.

No one can guarantee what the result of the suggested test would be. Because the SRT approach is free of inherent contradictions and resolves "the singularity problem" we expect that it would be in favor of the SRT approach with a break-through prospect in a gravitational theory development.

All sort of particles were somehow tested on a free fall in a gravitational field except a photon. Today's experimental technique allows testing it. One can go in long discussions of a current status of gravitational theories and their concepts of a photon. But it is hard to argue that a test of gravitational properties of a photon is of fundamental importance and should be done regardless of a personal preference of theory.

*Direct test a photon free fall in a gravitational field.*

As was shown, a photon emitted along a plane equipotential surface does not fall in a gravitational field as a massive particle does. This prediction may be checked experimentally. The experiment seems to be technically difficult but worth doing because it would be a direct check of photon gravitational properties.

*On a metric determination*

In the alternative approach a metric determination is principally different from that in the GRT. It is uniquely defined on the basis of the SRT mass-energy concept. The consequence of the concept is a frequency constancy of a photon in flight in a gravitational field. On the other hand, a proper mass and emission frequency of a chosen quantum-mechanical oscillator (atomic or nuclear resonant photon emitter-detector called *a standard clock*) as well as a photon speed, all these quantities are gauged by the factor $n_g(\phi)$, (the gravitational index of refraction). Consequently, a wavelength of the standard photon in a moment of emission does not depend on the gravitational potential. It gives a universal standard unit of length, which is necessary for construction of a global system of coordinates.

Let us consider for simplicity a spherically symmetrical field and a uniform gravitational field in a space within a thin spherical layer $\Delta x$ about a massive sphere. The layer is labeled by a potential $\phi(x)$, hence may be called $\phi$-world. Each $\phi$-observer may use his standard $\phi$-clock and compare its rate with observers in other $\phi$-worlds by exchanging a light signal. They will detect a frequency shift (a potential difference) and a corresponding variation of the speed of light. The procedure should be fast enough for making "quasi-stationary snaps" of the system. These unique features of matter allow mapping space-time in terms of Minkowski $\phi$-metric fully complying with the SRT mass-energy concept and Newtonian Physics in a limit.

### *On violation of the EP*

The question is whether a $\phi$-observer is able to detect "disappearing field" (or change of gauging factor $n_g(\phi)$) by any physical means without communication with other observers. We showed that the PRS experiment is the one of type, in which a small effect of a physical difference of two neighboring $\phi$-worlds is detected by means of information exchange, for example, by detecting a small difference in energy states of the emitter and detector. The effect is due to a proper mass change in a uniform field, that is, of the second order under weak field conditions. As was shown above (23), it is proportional to a potential difference between an emitter and a detector, placed on different equipotential surfaces, $\Delta m/m \approx g\Delta x/c^2$. For the effect of a proper mass variation is not affected by the state of free fall a violation of the EP is predicted in the suggested test. In fact, the test may be considered as probing a space with respect to an asymptotically flat space $\phi = 0$. At this point we come to the border of Gravitational Physics and Cosmology what is out of scope of this work.

In view of what is said above, one can regard the EP as a two-dimensional principle: it is valid on an equipotential surface (within a however thin layer) characterized by a flat quasi-stationary local metric. Freely falling frames should be considered equivalent inertial frames if and only if a change of the gauging factor $n_g(\phi)$ could be ignored (depending on the formulation of a physical experiment).

For an outside observer the whole space is seen as a "$\phi$-labeled" Minkowski space $\{x^\alpha, \tau(\phi)\}$ consisting of $\phi$- worlds sorted by $n_g(\phi)$. The factor determines an inner Minkowski $\phi$-metric characterized by the proper speed of light and the proper mass and frequency, all reduced by the factor $n_g(\phi)$ as compared to asymptotically flat space. Therefore, the outside observer investigating $\phi$-worlds is able to detect a traditional apparent effect of a gravitational time dilation (a red shift) as well as time delay effect due to decrease of the speed of light as measured by an outside observer. The physical meaning of those effects, of course, differs from that in the GRT. Quantitative difference rises with field strength.

*On mass and force categories*

As was shown, a relativistic gravitational and non-gravitational (inertial) masses differ in many ways, though in the Newtonian limit the difference disappears. This mass categorization is fundamentally important for the GRT because the EP was introduced through postulated equality of gravitational and inertial mass. The SRT mass-energy concept gives another categories of mass: a proper and kinetic mass. All above categories are interrelated but only the proper-kinetic mass concept allows understanding a formulation of Mach's question on an origin of inertial mass. Moreover, one can gain insight into the problem of unification of different types of forces through analysis of relativistic mass properties. Next illustrations show what kind of analysis we mean.

The proper mass of a freely falling particle characterizes a gravitational potential at a given point in space, so changing a proper mass and changing a gravitational potential is the same process expressed by different words. The kinetic mass is driven by a proper mass change while the total mass being kept constant. Therefore, a free particle has a maximal proper mass at infinity.

An action of any non-conservative (inertial) force on a test particle also causes a change of the proper mass. But now a potential difference arises due to work at expense of energy from external source. Therefore, a proper mass can increase and exceed that of a free particle at infinity.

Suppose two forces (conservative and non-conservative) compensate each other. Then there is no mass-energy current though a proper mass differs from that of a fee particle. For example, a repulsive electric force cannot emerge without work on like-charge particles to bring them at points with a given separation. As a result, a proper mass of a test particle may indefinitely increase. This is the way of a "man-made gravity" production.

A particle having increased proper mass and being released from an electric field acquires a kinetic energy in accordance with (18c) and eventually regains its "natural" proper mass. This SRT Dynamics is absent in Relativistic Quantum Electrodynamics.

*On a prospect of unification of electromagnetic and gravitational field*

The SRT approach casts a new light on the problem of an ultimate energy and dynamical properties of a material system. As is shown, a singularity of the point-like source turns out to be eliminated due to limited total mass-energy of a single particle. Consequently, a potential function deviates from the inverse distance law. Renormalization in Quantum Electrodynamics is not needed now. Both gravitational and electromagnetic forces seem to arise due to the same source of energy, that is, a proper mass of interacting particles. If so, the SRT mass-energy concept should be considered the basis for the next program of a unification of electromagnetism and gravity. The theory should give a consistent physical description of motion and radiation of a point-like neutral or charged massive particles in combined gravitational and electromagnetic field in the presence of external inertial forces.

One would have to deal with new gravitational properties of light such as the absence of its coupling to gravity and variation of the speed of light in a gravitational field. The problem is how to extend the Lagrangian formulation to the SRT Mechanics with inclusion of gravitational physics and find the unique field equations minimizing a system proper mass. Obviously, it would be a non-linear gauge theory ensuring the SRT mass-energy concept of energy-momentum conservation. Non-linearity in both gravitational and electromagnetic field would be due to proper mass dependence on a potential. The total energy should correspond to the total proper mass of a system when the system is decomposed into elementary particles moved to infinity. Energy sources providing inertial forces should be included in a total energy balance. Introduction of non-rigid bodies or linked particle systems would require a corresponding condensed matter model of atomic binding.

The SRT Mechanics equations discussed are not complete in this way. Though the equations (17a) and (17b) include the effect of a proper mass variation, still a variation of the speed of light is not introduced in an explicit form. The Lagrangian Relativistic Mechanics problem should be reformulated with specification of three types of forces (gravitational, electromagnetic, and non-conservative inertial) under conditions of variable proper mass and speed of light. It is a topic of another work.

*On a prospect of quantization*

Dealing with a relativistic continual variation of a proper mass one has to come up with the idea of a mass quantization and its connection to the De Broglie wave phenomenon. There must be an intimate relation between quantization of energy-momentum and a varying proper and kinetic mass.

*On universal physical constants*

A revision of the GRT if happened would most likely require a change of status of known universal physical constants. It was shown that in the alternative approach the speed of light, electric constant and masses of particles are functions of a gravitational potential. The quantities reach their maximal values in a "free space", the concept of which needs bridging gravitational and cosmological theories, on one side, gravitational and quantum-mechanical theories, on the other side.

*Some consequences*

Examples of consequences of the suggested approach in different physical branches are given below.

- *Super-luminous particles in cosmic rays.* High-energy cosmic particle when entering a region of a gravitational field, at some threshold energy become super-luminous. For example, a proton of energy about $5*10^{14}$ eV grazing the Sun exceeds the local speed of light. The threshold energy is proportional to the distance from the center of the Sun.

There could be some speculations about Physics of cosmic rays. The Sun may be considered a typical star in the universe, and cosmic rays could be a cosmological phenomenon. If so, particles in energy range from about $10^{15}$ up to $10^{18}$ eV while crossing regions of a size up to a thousand radii about a star become super-luminous. The particles will lose their energy due to Cherenkov radiation. Therefore, in the above energy region "a knee" giving a steeper slope in the spectrum should appear. This picture is in fact observed having no explanation. One cannot exclude that the clue given above may lead to a theoretical model reflecting physical reality.

- *Black hole phenomenon.* The GRT black hole concept is an extension of singularity of a point-like particle to the cosmic scale. The object should reach a critical magnitude of density and mass to become a black hole when gravitational and informational collapse occurs. Everything light included will be trapped. According to the GRT a time of a black hole formation is infinite. Because a universe lifetime is finite such objects could not exist but strangely enough they are allegedly observed.

In our approach there is no singularity, and objects with the above Properties do not exist in nature. Hence the above paradox does not take place. Instead there should be a smooth distribution of different massive objects with however huge binding energy. Light could always escape from them though red shift and time delay could be however great.

There is a guess on how so called Schwarzschild's (singularity) radius may appear in a theory. For a quadratic metric element we have a gauge factor $\exp(-2GM/c^2 x)$, which comes due to a proper mass variation and naturally eliminates singularity. In a weak field approximation it takes a form $1 - 2GM/c^2 x$. One can ignore a proper mass variation, take the latter for the general case, and get the radius.

- *Trouton and Noble experiment* [9]. This classical experiment has showed absence of a torque exerted on two charges at rest in a frame $S'$. The latter moves uniformly with respect to a frame $S$. The torque is expected due to a magnetic field generated by the two charges moving in the frame $S$. Therefore, the torque has a relativistic origin. The experimental result was interpreted as a proof of absence of the Ether.

In the following thought experiment let a repulsive electric force between two particles be compensated by a corresponding gravitational force. Then a torque due to each force should be equal and opposite *to comply with the Relativity Principle*. It means that a gravitational force is subject to the Lorentz transformation exactly in the same way as any force of other type. This prediction could be tested and undoubtedly confirmed.

### *Conclusion*

In this work we attempt to revise the foundation of the GRT, and explain reasons and idea. As always in such situations, there could be disbelief and arguments. That is why we suggest a falsifying test and emphasize its importance regardless of a theory status: the test is intended to reveal new gravitational properties of a photon. At the same time we think the explanation of why we predict "unexpected" result is also important.

Our arguments are based on the SRT mass-energy concept, which requires a dependence of a proper mass on a potential energy. It means that a freely falling particle has a varying proper mass. The author does not claim a discovery of the phenomenon: it was always a part of the SRT but barely considered in practice, and ignored in the GRT. If accounted for, a multitude of freely falling reference frames cannot be regarded as the multitude of equivalent (Einstein's) inertial reference frames. This is a starting point in the alternative approach to the development of a gravitational field theory on the SRT basis.

From the SRT mass-energy concept one may come to gravitational properties of matter different from those commonly thought. If true then an observer in a freely falling frame is able to identify his state by probing his local space with a photon. In other words, a violation of the EP is predicted in the suggested test ("weighing a photon in a freely falling frame").

The suggested test is crucial. In combination with the original PRS (stationary) experiment a new measurement of a frequency shift in a freely falling frame will allow discriminating between the two different approaches to the field theory. The question to be answered is whether a photon gravitates. The question has never been put in this form. We showed that existing observational data are not sufficient to make a conclusion in this respect. Luckily, the Mossbauer instrumentation is available, which can be used for further probing a gravitational field with a photon with a precision adequate to the test requirements.

### *References*

1. Clifford M. Will. Theory and Experiment in Gravitational Physics. Revised Edition. Cambridge, University Press (1993).
2. R.V. Pound and G.A.Rebka. Phys. Rev. Letters, **3**, 439 (1959), also:

   Phys. Rev. Letters, **4**, 337 (1960). R.V. Pound and J.L. Snider. Phys. Rev. Letters, **13**, 539 (1964).

3. J.L. Synge. Relativity: The Special Theory. North-Holland Publishing Company, Amsterdam (1965).

4. V.Fok. *The Theory of Space, Time, and Gravitation*. Translated by N.N. Kemmer (2d rev.ed.), Macmillan, New York (1964).

5. S.Weinberg. *Gravitation and Cosmology: Principles and Applications of the general Theory of Relativity*. John Willey @ Sons, Inc. New York (1972).

6. C.W. Misner, K.S. Thorne, J.A. Wheeler. *Gravitation*. W.H. Freeman and Company, San Francisco (1971).

7. A. Pais. *The Science and the Life of Albert Einstein*. Clarendon Press. Oxford, Oxford University Press. New York (1982).

8. C. Moller. *The Theory of Relativity*. (2d ed.). The International Series of Monographs on Physics. Oxford University Press (1982).

9. P. Lorrain, D. Corson. *Electromagnetic Fields and Waves*. Second Edition. W.H. Freeman and Company, San Francisco (1970). Also: Baldassare DiBartolo. *Classical Theory of Electromagnetism*. Prentice Hall, New Jersey (1991).

2929